 \documentclass[final,5p,times,twocolumn,nonatbib,authoryear]{elsarticle}

\usepackage{amssymb}
\usepackage{lipsum}
\usepackage{listings}
\usepackage{graphicx}
\usepackage{subcaption}
\usepackage{url}
\usepackage{multirow}
\usepackage{float}
\usepackage{rotating}
\usepackage{array}
\usepackage{enumitem}
\usepackage{amsmath}
\usepackage{xcolor}

\usepackage{fancyvrb}
\usepackage{tcolorbox}
\tcbuselibrary{breakable}
\usepackage{booktabs}
\lstdefinestyle{configstyle}{
    backgroundcolor=\color{gray!10},
    basicstyle=\ttfamily\footnotesize,
    keywordstyle=\color{blue}\bfseries,
    stringstyle=\color{orange!80!black},
    commentstyle=\color{gray}\itshape,
    breaklines=true,
    breakatwhitespace=true,
    frame=single,
    rulecolor=\color{gray!50},
    numbers=left,
    numberstyle=\tiny\color{gray},
    numbersep=5pt,
    tabsize=4,
    showstringspaces=false,
    captionpos=b,
}
\usepackage[natbibapa]{apacite}
\journal{Government Information Quarterly}

\begin{document}

\begin{frontmatter}

\title{Making Local Government Contracts Legible: A Computational Pipeline for Classifying and Mapping Intergovernmental Service Agreements}

\author[uvm]{Mohsen Ghasemizade\corref{cor1}\fnref{eq}}
\ead{Mohsen.Ghasemizade@uvm.edu}

\author[rutgers]{Raúl Gutiérrez-Meave\corref{cor1}\fnref{eq}}
\ead{raul.gmeave@rutgers.edu}

\author[vcsc]{Cailin Gramling}
\author[vcsc]{Aviral Chawla}
\author[vcsc]{Michael Robinette}
\author[uic]{Kate Albrecht}
\author[uvm,vcsc,csh]{Juniper Lovato}

\affiliation[uvm]{organization={University of Vermont, Department of Computer Science},
    city={Burlington},
    state={Vermont},
    country={United States}}

\affiliation[rutgers]{organization={Rutgers University-Newark, School of Public Affairs and Administration},
    city={Newark},
    state={New Jersey},
    country={United States}}

\affiliation[vcsc]{organization={University of Vermont, Vermont Complex Systems Institute},
    city={Burlington},
    state={Vermont},
    country={United States}}

\affiliation[uic]{organization={University of Illinois Chicago, Department of Public Policy, Management, and Analytics},
    city={Chicago},
    state={Illinois},
    country={United States}}

\affiliation[csh]{organization={Complexity Science Hub},
    city={Vienna},
    state={},
    country={Austria}}  

\cortext[cor1]{Corresponding author}
\fntext[eq]{These authors contributed equally}

\begin{abstract}
Interlocal agreements are one of the primary instruments through which local governments formalize collaboration for public service delivery, yet the institutional and financial content encoded in these contracts has remained inaccessible to systematic analysis at scale. This paper introduces an end-to-end computational pipeline for classifying intergovernmental agreements by institutional form and extracting financial relationships between principals and agents in service contracts. Applied to Iowa's 28E archive (N = 21,629), the largest dataset of interlocal agreements in the United States, the pipeline combines LLM-based summarization and classification across LLaMA 3.1, GPT 5.2 Pro, and Gemini 3 Pro on a four-class classification task that distinguishes agreements as either service contracts, resource sharing agreements, joint operations agreements, or new joint entity agreements. We also identify the financial principal and agent in these agreements and contracts, as well as the resulting dollar amounts and represent them on a directed network. The resulting financial network is organized around a small number of dominant service providers, with counties serving as the most structurally versatile actors, and cities as predominantly principals. By rendering the content of Iowa interlocal agreements analyzable at scale for the first time, this pipeline establishes a reusable methodology that researchers and state agencies can apply to track how public dollars move across local governments and to identify entities that depend heavily on a small number of providers.

\end{abstract}

\begin{graphicalabstract}
\includegraphics[width=\textwidth]{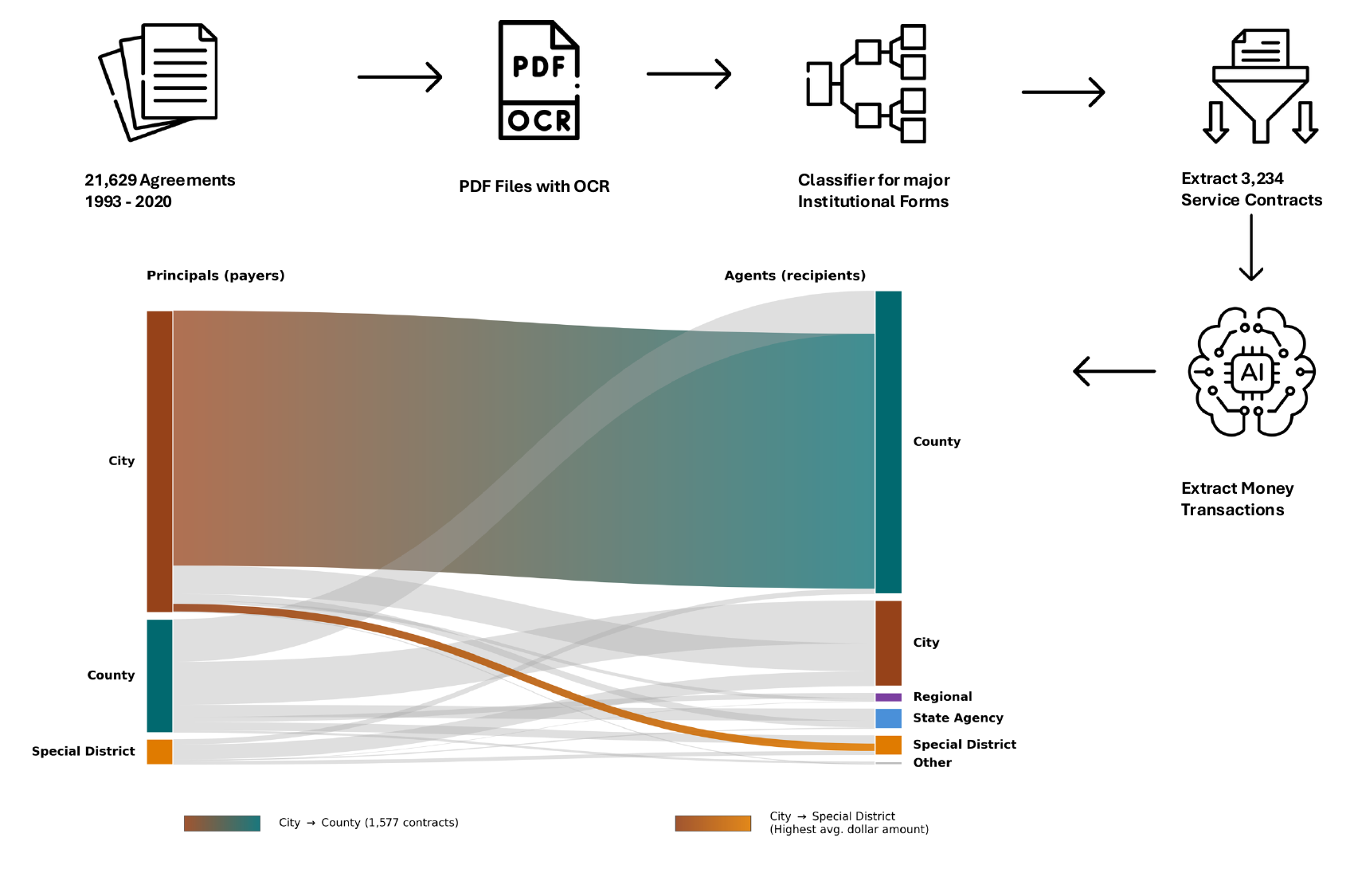}
\end{graphicalabstract}


\end{frontmatter}




\section{Introduction}

Interlocal agreements (ILAs) are one of the primary instruments through which local governments in the United States formalize collaboration for public service delivery \citep{andrew2009recent, thurmaier2002interlocal, leroux2010interlocal, gutierrez2025integrating, zeemering2008governing}. For example, cities contract with counties for police protection. Counties coordinate with neighboring counties on road maintenance and emergency management. School districts purchase services from municipalities. These arrangements are not incidental to local governance; they are an important means by which public services are organized and financed across fragmented jurisdictions \citep{hatley2014toward, thurmaier2016interlocal, leroux2010prospects}. 

Yet for all their importance, interlocal agreements have been remarkably difficult to study at scale. Filed as unstructured and noisy legal documents (structured as PDFs), they resist systematic analysis, and studies of interlocal collaboration have largely treated them as homogeneous (coding a tie between two governments as simply present or absent), collapsing institutional variation that has direct implications for how authority is shared, costs are divided, and accountability is structured \citep{li2021local, kurtz2026network, siciliano2021analyzing, sanchez2024little, gutierrez2025integrating, leroux2010interlocal}. This paper addresses that gap with a computational pipeline that makes the institutional and financial content of interlocal agreements analyzable at scale. The pipeline uses Optical Character Recognition (OCR) to extract text from scanned contracts,  Large Language Models (LLMs) to classify each agreement's institutional form and identify who pays whom and how much, and network analysis to reveal the resulting structural patterns. Applied here to Iowa's 28E archive, the pipeline is not specific to this dataset: the same approach can be applied to other states' interlocal agreements, and more broadly to other large collections of legal or administrative contracts. A detailed schematic of the full pipeline, including tools, model versions, parameters, and the data representation at each stage, is provided in \ref{app:pipeline} (Figure~\ref{fig:pipeline_flowchart}).

Variation in this analysis matters for understanding the diverse ecosystem of agreement types and the various players involved in making agreements. The institutional mechanism governments choose to provide services, whether a resource sharing arrangement, a service contract, a joint operation, or a new joint entity, reflects the depth of their commitment, the risks they are willing to absorb, and the governance demands they are prepared to meet \citep{thurmaier2002interlocal, kim2022updating, carr2013costs}. For example, a service contract establishes a principal-agent relationship in which the principal pays for the service and the agent delivers it, with financial obligations and defined accountability structures. Joint operations establish partnerships to co-produce services on an ongoing basis. New joint entities represent the most integrated form, in which participating governments establish a legally distinct body to manage or deliver a shared service, with its own governance dynamics \citep{morse2014mapping}. Without a pipeline that distinguishes these forms, a large-scale dataset would treat all of them as the same kind of arrangement. Treating these forms as equivalent obscures the institutional architecture of local public service delivery and limits what researchers and practitioners can learn from the agreements governments file \citep{carr2024functional}.

The state of Iowa offers a rare opportunity to study institutional variation in interlocal agreements at scale. Under Chapter 28E of the Iowa Code, local governments are legally required to file all collaborative agreements with the Secretary of State before they take any effect \citep{contracts}. No other state imposes this mandate, and the result is a uniquely comprehensive public archive: 21,629 agreements filed between 1993 and 2020, covering every type of local government across thirty-three service areas \citep{contracts}. The archive is not only large, but it is also complete in a way that no other state-level dataset of interlocal agreements can match, making it the appropriate foundation for studying how local governments structure collaboration and finance service delivery at scale.

The challenge is that completeness does not equal accessibility. The agreements are stored as PDF files, many of them scanned images of signed paper documents, which has confined most prior analysis to small scale, manual coding.
A Governance Lab in a University in The U.S. has hand-coded 1,128 of these agreements and hand-classified their institutional form, a process that is both time-consuming and difficult to scale. Automatically extracting the institutional and financial content encoded in the contracts requires solving a set of computational problems that off-the-shelf tools are not designed to handle, such as noise from PDF documents, noise from OCR processing, class imbalance across institutional forms, document lengths that exceed standard model context windows, and a labeled training set small enough to make standard supervised learning unreliable. LLMs are a promising step forward for this kind of complex information-extraction task, but their application in specialized, noisy, legal domains requires careful methodological design. \cite{dahl2024large} find that LLMs hallucinate in legal contexts at a rate of 58\%, and that models struggle to identify even their own errors, making validation an essential component of any pipeline applied to government contracts.

This paper introduces an end-to-end computational pipeline that addresses these challenges directly. The pipeline combines state-of-the-art OCR processing, LLM-based summarization and classification, embedding-based machine learning, and semi-supervised learning to classify the full Iowa 28E archive by institutional form and extract the financial structure of service contracts. Applied to Iowa's archive, it produces the first large-scale classified dataset of interlocal agreements in the United States and uses it to map how money moves across local governments through formal service arrangements. The findings reveal a system organized around a small number of dominant service providers, where cities are predominantly buyers, counties are structurally central as both buyers and sellers, and special districts occupy a dependent position at the margins of the contracting network.

Because these barriers made large-scale, automated extraction of contractual detail infeasible until recently, prior network analyses of interlocal agreements using these data have treated ties as undirected, representing collaboration as a symmetric relationship between partners (see, for instance, \citet{carrillo2023ebb, kurtz2026network,gutierrez2025integrating, meave2026sets, li2021local, siciliano2021analyzing, sanchez2024little}). Service contracts, however, are not symmetric. One government pays, and one government delivers, and that asymmetry is the substance of the accountability relationship encoded in the contract. This paper recovers that directionality at scale for the first time, constructing a directed financial network in which edges flow from principal to agent and are weighted by contract value.

Our primary contributions are threefold. First, we introduce a computational pipeline and classification schema for intergovernmental agreements that can be applied at scale, demonstrating its use on a dataset of over 21,000 agreements and making the pipeline code and detailed methodology publicly available to enable systematic research and institutional oversight. Second, we present a reproducible approach that combines LLM-based classification with traditional machine learning to overcome the core challenges of working with unstructured legal documents at scale, including noise, class imbalance, document length, and small labeled datasets. Third, we provide a case study of financial relationships between principals and agents in Iowa service contracts, mapping the network structure of these transactions and analyzing the dyadic and structural patterns that characterize how local governments finance public service delivery, offering insights relevant to computational researchers, legal scholars, and scholars of local governance.

\section{Related Work and Background}

\subsection{Interlocal Collaboration Agreements and Institutional Form}

Interlocal agreements are written contracts through which local governments formalize collaboration for public service delivery \citep{andrew2009recent, thurmaier2002interlocal, leroux2010interlocal, zeemering2008governing}. They are a primary mechanism through which fragmented local government systems coordinate across jurisdictional boundaries, enabling municipalities, counties, townships, school districts, and special districts to share resources, deliver services, and build governance arrangements that no single government could sustain independently \citep{carr2016city}.

These agreements vary considerably in how they structure collaboration. \cite{morse2014mapping} propose a continuum of institutional forms ranging from minimal integration arrangements to fully merged departments, including resource sharing arrangements in which governments exchange or jointly use information, personnel, equipment, or facilities; service contracts in which one party provides a service to another under defined financial or operational terms; joint operations that establish partnerships to co-produce services on an ongoing basis; and new joint entities in which participating governments establish a legally distinct body to manage or deliver a shared service. Collaboration is therefore not a binary choice but a structured decision about how much institutional integration governments are willing to commit to, and the form chosen carries consequences that extend well beyond the initial agreement.

The choice among these forms is not arbitrary. Drawing on institutional collective action theory, governments select institutional arrangements that balance the benefits of collaboration against the transaction costs of collaborating across organizational and jurisdictional boundaries \citep{feiock2013institutional, kim2022updating}. More integrated forms offer greater capacity for sustained collaboration but require governments to absorb higher governance demands, including the costs of monitoring, enforcement, and managing interdependence over time. The risks of service collaborations vary systematically with the institutional form chosen, and governments that underestimate them face significant challenges in sustaining collaborative arrangements \citep{carr2013costs}. Institutional form is therefore not a procedural detail but a governance choice with direct implications for authority allocation, cost division, and accountability structure across participating governments.

Among all institutional forms, service contracts make accountability stakes most visible. As the most common form in Iowa's 28E agreements \citep{li2021local}, they establish a financial principal-agent relationship in which one government pays for a service, and another delivers it under defined terms. Principal-agent theory identifies this as a fundamental governance problem: the principal must design contractual arrangements that align the agent's behavior with its own objectives, while managing the informational asymmetries that make monitoring costly \citep{gailmard2014accountability, braun2003principal}. In interlocal service contracts, the obligations of each party are encoded directly in the agreement, making the contract itself the primary instrument through which accountability is defined and enforced \citep{van2024contracts, zhang2024exploring}. Which governments occupy principal roles, which serve as agents, and how financial resources flow across these relationships are governance questions central to understanding local service delivery, yet they remain largely unanswered at scale due to the inaccessibility of contract-level data.

\subsection{From Documents to Data: Computational Methods for Legal and Government Text}

Government and legal documents can vary from brief paragraphs to lengthy multiple page agreements, requiring different strategies for classification. Researchers have developed hierarchical neural networks that mirror document structure for long texts. \cite{yang2016hierarchical} encoded words into sentence representations, then sentences into a document representation, with attention mechanisms at each level to focus on the most informative content. Such hierarchical-aware classifiers have significantly outperformed flat models on long document classification tasks. Recent transformer-based methods address the challenge of lengthy documents that exceed the typical 512-token limit of BERT~\citep{devlin2019bert}. Specialized models like Longformer \citep{beltagy2020longformer} and BigBird \citep{zaheer2020big} use sparse and global attention patterns to scale to thousands of tokens \citep{dai2022revisiting}. 

In many scenarios, only short textual fields need classification, for example, contract titles, clause headers, or brief descriptions. For such texts, simple but tedious representations often suffice. Studies in text classification have shown that a bag-of-words or n-gram representation with a linear classifier remains a competitive baseline for short inputs \citep{yang2016hierarchical}. In addition, character-level convolutional neural network (CNN) models can excel in short government text that contains abbreviations or typos \citep{muir2021using}.

\subsubsection{Noisy Data and Class Imbalance}

Public sector data often contains noise like misspellings, abbreviations, or errors. Such noise can seriously degrade classification performance. For example, a study of BERT found that the introduction of common typos and spelling mistakes decreases accuracy in sentiment detection tasks by around 8\% \citep{kumar2020noisy}. This highlights the need for robustness when dealing with government records and for using a state-of-the-art OCR tool, such as SURYA~\citep{paruchuri2025surya}, in this work to reduce noise from scanned documents.

To make the classifiers more resilient to noise, one approach is to simulate noise in training data \citep{xu2021robust}, such as synthetic OCR noisy data, to improve the model's tolerance to transcription mistakes. Other techniques include noise-tolerant loss functions and data analysis with spelling variants. Empirically, these strategies can substantially boost the robustness of pre-trained models to noisy inputs.

Another common challenge in legal datasets is imbalanced classes, where some agreement types or outcomes are very rare. Imbalanced data skews classifiers toward the majority class. \cite{chawla2002smote} formally highlighted this issue, noting that many real datasets consist of a large portion of `normal' instances and only a small percentage of `interesting' or `minority' cases. Misclassifying those rare but important cases has high cost. To address this, resampling techniques are widely used. The SMOTE algorithm produced by \cite{chawla2002smote} creates synthetic minority examples and combines this oversampling with undersampling of the majority class. More recent advances leverage large pretrained language models to generate fully synthetic in-domain texts: by prompting models like LLaMA or GPT with a few exemplar summaries for an underrepresented class, one can produce novel contract descriptions that enrich the minority class. But this method lacks diversity, and it is hard to validate the quality and accuracy of the synthetic data.

\subsubsection{Large Language Models in Classification}

The advent of large pre-trained language models has greatly advanced text classification. \cite{devlin2019bert} introduced BERT, a bidirectional encoder transformer trained on massive corpora with 340 million parameters, which can be fine-tuned for downstream tasks by adding a simple output layer to benchmark text classification across different topics. Building on BERT, researchers have crafted domain-specific language models to further improve performance on specialized text. RoBERTa, an improved version of BERT trained on a larger dataset, has shown good accuracy across different domains with sufficiently large labeled datasets, such as the 87\% accuracy reported for a 1,800 example conspiracy theory classification task \citep{ghasemizade2024developing}. Legal-BERT \citep{chalkidis2020legal}, pre-trained on large legal corpora, provides vocabulary and representations tailored to legal jargon.

Meanwhile, massive language models such as OpenAI's GPT-5 have shown that simply scaling model size can unlock new classification capabilities. They have suggested that with sufficient pre-training, a model can perform tasks such as classification in few-shot or zero-shot settings. More recently, even larger and more capable `reasoning' models, designed to generate contextual embeddings and explicit chains of thought, have become available. OpenAI's GPT-5.2 \citep{singh2025openai} exhibits stronger zero and few-shot classification performance on legal text. Similarly, Google's Gemini models, particularly Gemini 3 Pro \citep{gemini3pro2025}, have demonstrated powerful reasoning and long-context understanding, establishing a new benchmark in multimodal, general-purpose AI. At the same time, open-access variants like Meta's Llama 3.1 \citep{grattafiori2024llama} offer comparable embedding quality in a smaller size, enabling us to extract sentence representations without the API limitations of closed systems. 

These recent computational techniques have enabled researchers across many fields to develop classification and regression pipelines to address specific research questions. Most of the machine learning models require a massive training dataset, and sometimes that dataset does not exist. That is when LLMs become particularly useful, because these are vast complicated models trained on complex datasets which can capture the diversity and complexity seen in these documents. Chain-of-thoughts reasoning, for example, gives us a reasonable classification accuracy with just one provided example. This makes it suitable for this task, where we have a small training size and noisy, long, and complicated legal documents.

\section{Methodology and Dataset}

Figure~\ref{fig:pipeline_flowchart} in~\ref{app:pipeline} provides a complete schematic of the pipeline described in this section, following standard flowchart conventions: each processing step, decision point, manual validation checkpoint, and intermediate data artifact is shown with the specific tool, model, and parameters used.

\subsection{Data and Corpus}
\subsubsection{Data Acquisition and Scope}

The dataset analyzed in this study is drawn from the Iowa Secretary of State's public archive of 28E intergovernmental agreements \citep{contracts}. Under Chapter 28E of the Iowa Code, local governments are legally required to file collaborative agreements with the Secretary of State in order for these contracts to be valid and enforceable. This statutory mandate creates a uniquely comprehensive record of formalized collaboration across municipalities, counties, townships, school districts, and special districts, as well as collaboration with non-profits and private entities.

The dataset was collected, cleaned, and organized by
a Governance Lab in the U.S. It covers 21,629 agreements filed between 1993 and 2020, spanning thirty-three service areas as defined by the Secretary of State. These include core categories such as police protection, fire response, education, and street and road systems, as well as more specialized areas like criminal investigation, jail and corrections, emergency management, water systems, economic development, and information services.

The rationale for focusing on Iowa's 28E agreements rests on two considerations. First, the statutory filing requirement ensures near-universal coverage of formal interlocal contracts within a state, reducing concerns about selection bias or incomplete records that often complicate research on collaboration. Second, the filing process yields consistently formatted filings, but not machine-readable ones, that enable large-scale text analysis and computational modeling. No other state imposes this mandate, making Iowa's archive the most complete basis available for studying how local governments structure collaboration and finance service delivery at scale. The challenge, however, is that completeness does not equal accessibility. The agreements are stored as PDF files, many of them scanned images of signed paper documents with stamps, and extracting their institutional and financial content at scale requires a computational pipeline designed specifically for the noise, variability, and complexity of government legal documents.

\subsubsection{Dataset Annotation and Labeling}

A total of 1,128 agreements were hand-coded by
a Governance Lab in the U.S. Based on the work by \cite{morse2014mapping}, the coding scheme distinguishes among the four institutional forms that capture the most common mechanisms through which local governments in Iowa structure collaboration. \textit{Resource sharing} refers to agreements in which governments exchange or jointly use information, personnel, equipment, or facilities. \textit{Service contracts} establish a relationship in which one party provides a service to another under defined financial or operational terms. \textit{Joint operations} establish partnerships to co-produce services on an ongoing basis. \textit{New joint entities} represent the most integrated form, in which participating governments establish a legally distinct body to manage or deliver a shared service. Each form carries different implications for how governments share risk, allocate authority and resources, and sustain collaboration over time \citep{carr2013costs, kim2022updating}. Table~\ref{tab:doc_stats} presents the distribution of contracts across institutional types and the average document length for each category.

\begin{table}[h]
\centering
\caption{Document statistics for the hand-labeled subset of Iowa 28E agreements, by institutional form ($N = 1{,}128$). \# Docs is the number of labeled contracts per category. Length counts number of words in each document.}
\begin{tabular}{lcc}
\hline
\textbf{Institutional Form} & \textbf{\# Docs} & \textbf{Mean Length} \\
\hline
Service Contract   & 589 & 12.4k \\
Joint Operations   & 250 & 17.7k \\
New Joint Entities & 157 & 16.2k \\
Resource Sharing   & 132 & 11.5k \\
\hline
\end{tabular}
\label{tab:doc_stats}
\end{table}

Hand-coding was conducted in two stages. First, two coders independently reviewed a subset of agreements to refine the annotation protocol and resolve ambiguities, producing a shared set of decision rules applied to the full sample. In the second stage, the remaining agreements were coded independently using the finalized protocol. The process produced high inter-coder reliability, with agreement rates above 95\% in the pilot stage and 99\% once the guidelines were stabilized. Any disagreements were resolved through discussion and group consensus.

\subsubsection{\textbf{Document Processing and OCR:}}

The complete dataset was stored as PDF files. Notably, some older agreements existed only as scanned images of physical, signed contracts rather than as digitally created documents. This distinction matters because scanned image-based PDFs cannot be read directly by standard text extraction tools, the document appears as a photograph rather than as searchable text.
To convert these scanned files into machine-readable text suitable for analysis, we applied OCR, a technology that interprets printed characters in images and transcribes them into digital text. We used two state-of-the-art OCR tools: General OCR Theory (GOT) \citep{wei2024general} and SURYA \citep{paruchuri2025surya}. GOT handles complex document layouts by recognizing, standard prose, tables and structured data, content types common in government contracts. SURYA complements this by first identifying the structural regions of each page (such as headings, body text, and tables) before applying a text recognition model, making it well-suited to the formatting conventions of public sector records. We select SURYA's output to feed our pipeline.

In our case, the complexity comes in the tables, signatures, boxes, stamps, and lines to be filled in the contracts. We employ both of these modern OCR tools to extract text documents from the PDF files and to report the character-error rate (CER) for each one. We randomly sampled a number of the PDF files and manually extracted the text from them as reference documents. To address the potential bias associated with selecting 100 samples, we utilize the confidence interval (\ref{sec:ocr}) to compare the selected OCR tools.

\subsection{LLM Based Classification}

\textbf{Stage 1: Summarization:}

One of the challenges of these agreements is the length, because not every architecture and algorithm can cover the whole document in one context window. Also, each contract has unrelated and unnecessary details for this classification task, which can mislead our classifiers. To overcome these challenges, we decided to summarize each contract by extracting the important parts, key entities, obligations, and the overall purpose. We used the aforementioned LLMs with big enough context window. Feeding the summaries to our next machine learning models, is also going be less computationally heavy.

To summarize the documents, we used this prompt across all models:

\begin{tcolorbox}[
    breakable,
    colback=gray!8,
    colframe=gray!50,
    title={\small\textbf{Contract Summarization Prompt}},
    fontupper=\ttfamily\footnotesize
]
Summarize the following contract text in 3--5 sentences.
Keep key entities, obligations, and the overall purpose.

CONTRACT TEXT:
[CONTRACT TEXT]
\end{tcolorbox}

\textbf{Stage 2: Chain of Thoughts:} 

Chain-of-thought (CoT) prompting instructs the model to work through a task as a sequence of explicit intermediate reasoning steps rather than producing an answer directly \citep{wei2022chain}. This improves accuracy on complex tasks, at the cost of substantially more computation per document. This method breaks down the complex problem into smaller, logical parts, and has shown improvements in the accuracy. The downside of this approach is the need for LLMs which have large parameter sizes (on order of billions) and this makes the task computationally intensive. For this work, we picked Llama 3.1 70B, OpenAI GPT 5.2 Pro, and Google Gemini 3 Pro to perform  zero-shot, 1-shot, and 3-shot learning. Zero-shot learning is when a model performs a task given only an instruction and no examples, while few-shot learning gives it a small handful of worked examples in the prompt to demonstrate the task before it attempts the real one. Due to the limitation of the context window, we only used summaries for this task. Note that, each contract has a short paragraph about the purpose of the contract, we added that next to our summary.

The example prompt for 1-shot learning:

\begin{tcolorbox}[
    breakable,
    colback=gray!8,
    colframe=gray!50,
    title={\small\textbf{Classification Prompt Template (1-shot)}},
    fontupper=\ttfamily\footnotesize
]
You are a contract assistant performing a classification task.
Please perform chain-of-thought reasoning to classify the
summary of the contract below.

Contract Purpose: [PURPOSE]
Contract Text: [CONTRACT SUMMARY]

Category Definitions \& Examples:

- Joint Operations: Organizations work together to jointly
  produce or deliver services.
  Example: [EXAMPLE SUMMARY]

- New Joint Entities: Organizations collaborate to create a
  new, legally distinct shared entity.
  Example: [EXAMPLE SUMMARY]

- Resource Sharing: Organizations share information,
  personnel, or physical equipment.
  Example: [EXAMPLE SUMMARY]

- Service Contract: One entity hires an outside public or
  private provider for specific services.
  Example: [EXAMPLE SUMMARY]

Return only a JSON object:
  \{``Target'': ``<one of the categories above>''\}.
\end{tcolorbox}

\textbf{Stage 3: Embedding-Based Classification:}

Machine learning models require text to be represented as numerical values before analysis can occur. Earlier approaches to this conversion, such as TF-IDF \citep{salton1988term}, Word2Vec \citep{mikolov2013efficient}, and GloVe \citep{pennington2014glove}, treated words largely in isolation, without accounting for how meaning shifts depending on context. More recent methods draw on LLMs to produce what are known as contextualized embeddings: numerical representations that capture the broader meaning of a passage in context, not just its individual words \citep{liu2024llmembed}. Research in legal and governmental text analysis suggests these richer representations outperform traditional approaches \citep{10825997}.

In this study, we extract embeddings from the contract summaries generated by our LLMs and use them as inputs to a range of classification models. These include both classic machine learning approaches, logistic regression and Support Vector Machines (SVM), and more computationally intensive methods: Extreme Gradient Boosting (XGBoost), and Multi-Layer Perceptron (MLP). This breadth allows us to assess whether the performance gains from contextualized embeddings hold consistently across model types, or whether they are specific to particular architectures.

\subsection{Network Construction}

Service contracts are the only institutional form in the corpus with a clear financial principal-agent structure: one government pays for a service and another delivers it under defined contractual terms. This asymmetry makes them the appropriate unit for mapping how resources flow across local governments, and we construct a directed financial network from them in which edges run from principal to agent and are weighted by contract value.

To construct the financial network, we use GPT-5.2 Pro to extract the contracting parties (the principal and agent) and the financial value of each agreement from every identified service contract, with the prompt below. Where an agreement specified recurring payments rather than a single payment, the model summed them into a total contract value. The analysis proceeds in three phases. The first addresses entity preparation: agency names are cleaned, disambiguated, and standardized across contracts before any network is constructed, since name variation across documents would otherwise fragment single actors into multiple nodes and distort every downstream measure. The second phase uses a configuration model null to assess which structural properties are substantively surprising rather than artifacts of the degree sequence, comparing observed network properties against 1,000 random rewirings that preserve in- and out-degree for every node. The third phase characterizes the basic structural properties of the contracting network, including density, reciprocity, transitivity, and centralization, to establish how financial relationships are distributed across government types and service areas, and applies a weighted hierarchical Stochastic Block Model \citep{peixoto2014hierarchical} in Graph-tool \citep{peixoto_graph-tool_2014} to recover the latent block structure of the contracting system without imposing a fixed number of groups. 

\begin{tcolorbox}[
    breakable,
    colback=gray!8,
    colframe=gray!50,
    title={\small\textbf{Financial Network Extraction Prompt}},
    fontupper=\ttfamily\footnotesize
]
You are a government contract analyst specializing in Iowa 28E interlocal agreements.

Your task is to extract structured financial and metadata fields from raw contract text.

Key rules:

- PRINCIPAL = the government entity that initiates and PAYS for the service.
  Sometimes the paying party is listed second; read the full text carefully.
  
- AGENT = the entity that PERFORMS the service and RECEIVES payment.

- Use full formal names (e.g., "City of Des Moines" not "Des Moines" or "the City").

- If no monetary transaction exists, set amount to null and has\_transaction to false.

- For agency types return exactly one of: city, county, school district, state agency, other government, private.
\end{tcolorbox}

\subsection{Entity Disambiguation and Node Construction}

Before building any network, agency names must be cleaned and standardized. The same agency appears under many different name variants across contracts (e.g., City of Des Moines', Des Moines City', `Des Moines, IA'). Duplicate nodes will silently break edge structure and bias every downstream analysis. We used fuzzy string matching via RapidFuzz \citep{max_bachmann_2024_10938887} across all extracted agency names to identify likely duplicates, built a canonical name list from the matches, and manually verified the top 100 most frequently appearing agencies. These are the high-degree nodes that drive most findings. Using a similarity threshold of 87, we merged 349 matches, yielding 1,642 unique canonical nodes.

\section{Pipeline Performance}

\subsection{OCR improvements}
\label{sec:ocr}

To measure the performance of our selected modern OCR tools, GOT and SURYA, in our dataset, we use a metric called Character Error Rate (CER). This metric counts the number of substitutions (S), deletions (D) and insertions (I) to achieve the reference output. The lower the value the better the performance of the OCR. 

$$CER = \frac{S + D + I}{S + D + C}$$

We manually constructed 100 of the random documents as the reference models and calculated the CER  using this framework on the HuggingFace platform \citep{morris2004} overall for all of them for both GOT and SURYA. GOT achieved an CER score of \textbf{20\%}, while SURYA showed significant improvement of \textbf{8.7\%}.

To verify that the observed 11.3\% performance gap is statistically significant and not a result of random chance, we calculated the 95\% confidence interval for the difference between these two error rates. The formula for the confidence interval of a difference in proportions is:

\[
\small
\text{CI}_{\text{diff}} = (\hat{p}_{\text{GOT}} - \hat{p}_{\text{SURYA}}) 
\pm Z \sqrt{\tfrac{\hat{p}_{\text{GOT}}(1-\hat{p}_{\text{GOT}})}{n_{\text{GOT}}} 
+ \tfrac{\hat{p}_{\text{SURYA}}(1-\hat{p}_{\text{SURYA}})}{n_{\text{SURYA}}}}
\]

\begin{align}
\text{CI}_{\text{diff}}
&= (0.20 - 0.087) \pm 1.96 \sqrt{\tfrac{0.20(0.80)}{100} + \tfrac{0.087(0.913)}{100}} \\[6pt]
&= 0.113 \pm 0.094 = (0.019,\; 0.207)
\end{align}

\subsection{Chain of Thought}

The results of the high-cost chain-of-thought classification task are presented in Table~\ref{tab:fewshot_results}. The GPT 5.2 Pro demonstrates the highest performance, achieving a weighted average F1 score of 0.74 with 1-shot learning. This result implies that even with a single example of the class, we can classify intricate and noisy legal documents with a reasonable score. This is a significant advancement, as it helps labeling datasets, provided that the achieved accuracy is deemed reasonable for the specific task.

\begin{table*}[ht]
\centering
\caption{F1-scores for chain-of-thought prompt-based classification, by number of in-context examples (0-shot, 1-shot, 3-shot), for three large language models: LLaMA, GPT, and Gemini. Models were prompted directly on the classification task with no downstream classifier trained; F1 is computed per institutional-form category against the same test held-out set of hand-labeled contracts. Macro Avg is the unweighted mean across the four classes; Weighted Avg weights each class by its number of test instances.}
\label{tab:fewshot_results}
\renewcommand{\arraystretch}{1.2}
\resizebox{\textwidth}{!}{
\begin{tabular}{l|ccc|ccc|ccc}
\hline
\multirow{2}{*}{\textbf{Class}}
  & \multicolumn{3}{c|}{\textbf{LLaMA}} 
  & \multicolumn{3}{c|}{\textbf{GPT}} 
  & \multicolumn{3}{c}{\textbf{Gemini}} \\
\cline{2-10}
  & \textbf{0-shot} & \textbf{1-shot} & \textbf{3-shot} 
  & \textbf{0-shot} & \textbf{1-shot} & \textbf{3-shot} 
  & \textbf{0-shot} & \textbf{1-shot} & \textbf{3-shot} \\
\hline
Joint Operations       & 0.50 & 0.57 & 0.58 & 0.59 & 0.61 & 0.60 & 0.50 & 0.51 & 0.51 \\
New Joint Entities     & 0.57 & 0.73 & 0.80 & 0.88 & 0.87 & 0.86 & 0.59 & 0.79 & 0.79 \\
Resource Sharing       & 0.48 & 0.63 & 0.64 & 0.60 & 0.63 & 0.60 & 0.33 & 0.43 & 0.29 \\
Service Contract       & 0.81 & 0.79 & 0.76 & 0.77 & 0.80 & 0.78 & 0.79 & 0.80 & 0.77 \\
\hline
Macro Avg              & 0.59 & 0.68 & 0.70 & 0.71 & 0.73 & 0.71 & 0.55 & 0.63 & 0.59 \\
Weighted Avg           & 0.67 & 0.71 & 0.71 & 0.72 & 0.74 & 0.72 & 0.67 & 0.68 & 0.65 \\
\hline
\end{tabular}
}
\end{table*}

\subsection{Embedding-Based Classification}
\label{sec:classifier}

We evaluated several classification algorithms on the embeddings extracted from each LLM, including logistic regression, support vector machines, ridge classifiers, ensemble tree methods, and multi-layer perceptrons, on an 80\% train, 20\% test split. All models were assessed using five fold cross validation with balanced class weights to account for the uneven distribution of contract types in our dataset. To ensure a fair comparison, we performed a systematic search over standard hyperparameter ranges for each algorithm and report results from the best performing configuration. Full implementation details and the hyperparameter search space are provided in~\ref{app:training_details}.

The strongest classifier paired with GPT embeddings was a support vector machine with a radial basis function kernel, achieving an overall accuracy of 82\% and a weighted F1-score of 0.82. As shown in Table~\ref{tab:gpt_rbfsvm}, performance was highest for `service contracts' (F1 =0.92), the most common category in the dataset, and lowest for `resource sharing' (F1 =0.61), which appear less frequently and often overlap conceptually with joint operations arrangements. We use this model for our pipeline.

\begin{table}[H]
\centering
\caption{Classification performance of a support vector machine with a radial basis function (RBF) kernel, trained on GPT-derived text embeddings. Each row reports precision (share of predicted-class labels that were correct), recall (share of true-class instances correctly identified), and F1 (the harmonic mean of precision and recall) for one of the four institutional-form categories. Accuracy is the share of all test documents classified correctly. Macro Avg is the unweighted mean across the four classes; Weighted Avg weights each class by its number of test instances.}
\label{tab:gpt_rbfsvm}
\renewcommand{\arraystretch}{1.2}
\begin{tabular}{lccc}
\hline
\textbf{Class} & \textbf{Precision} & \textbf{Recall} & \textbf{F1} \\
\hline
Joint Operations    & 0.59 & 0.74 & 0.65 \\
New Joint Entities  & 0.92 & 0.54 & 0.68 \\
Resource Sharing    & 0.87 & 0.46 & 0.61 \\
Service Contract    & 0.89 & 0.95 & 0.92 \\
\hline
Accuracy            &      &      & 0.82 \\
Macro Avg           & 0.82 & 0.68 & 0.72 \\
Weighted Avg        & 0.83 & 0.82 & 0.82 \\
\hline
\end{tabular}
\end{table}

For LLaMA embeddings, logistic regression yielded the best results, with an accuracy of 77\% and a weighted F1-score of 0.77 (Table~\ref{tab:llama_logreg}). The pattern across classes mirrored that of GPT: strong performance on service contracts (F1 =0.88) and weaker performance on resource sharing (F1 =0.47). Notably, this embedding based approach offered only modest gains over direct chain-of-thought prompting of the same LLaMA model, which reached a weighted F1-score of 0.71. 

\begin{table}[H]
\centering
\caption{Classification performance of a logistic regression model trained on LLaMA-derived text embeddings. Each row reports precision (share of predicted-class labels that were correct), recall (share of true-class instances correctly identified), and F1 (the harmonic mean of precision and recall) for one of the four institutional-form categories. Accuracy is the share of all test documents classified correctly. Macro Avg is the unweighted mean across the four classes; Weighted Avg weights each class by its number of test instances.}
\label{tab:llama_logreg}
\renewcommand{\arraystretch}{1.2}
\begin{tabular}{lccc}
\hline
\textbf{Class} & \textbf{Precision} & \textbf{Recall} & \textbf{F1} \\
\hline
Joint Operations    & 0.51 & 0.63 & 0.56 \\
New Joint Entities  & 0.84 & 0.67 & 0.74 \\
Resource Sharing    & 0.54 & 0.40 & 0.46 \\
Service Contract    & 0.87 & 0.88 & 0.88 \\
\hline
Accuracy            &      &      & 0.77 \\
Macro Avg           & 0.70 & 0.64 & 0.66 \\
Weighted Avg        & 0.77 & 0.77 & 0.77 \\
\hline
\end{tabular}
\end{table}

For Gemini embeddings, a polynomial-kernel support vector machine yielded the best performance, with an overall accuracy of 77\% and a weighted F1-score of 0.76 (Table~\ref{tab:gemini_polysvm}). The familiar pattern across classes held: `service contracts' were classified most reliably (F1 =0.83), while `joint operations' (F1 =0.62) and resource sharing (F1 =0.65) remained the more difficult categories. New joint entities, despite being the second rarest class in the test set, were identified with notably high precision (0.92), suggesting that when Gemini embeddings flag an agreement as belonging to this category, that label is unusually trustworthy, a useful property for downstream review workflows where false positives are costly.

\begin{table}[H]
\centering
\caption{Classification performance of a support vector machine with a polynomial kernel, trained on Gemini-derived text embeddings. Each row reports precision (share of predicted-class labels that were correct), recall (share of true-class instances correctly identified), and F1 (the harmonic mean of precision and recall) for one of the four institutional-form categories. Accuracy is the share of all test documents classified correctly. Macro Avg is the unweighted mean across the four classes; Weighted Avg weights each class by its number of test instances.}
\label{tab:gemini_polysvm}
\renewcommand{\arraystretch}{1.2}
\begin{tabular}{lccc}
\hline
\textbf{Class} & \textbf{Precision} & \textbf{Recall} & \textbf{F1} \\
\hline
Joint Operations    & 0.64 & 0.60 & 0.62 \\
New Joint Entities  & 0.92 & 0.85 & 0.88 \\
Resource Sharing    & 0.64 & 0.66 & 0.65 \\
Service Contract    & 0.82 & 0.85 & 0.83 \\
\hline
Accuracy            &      &      & 0.77 \\
Macro Avg           & 0.75 & 0.74 & 0.74 \\
Weighted Avg        & 0.76 & 0.77 & 0.76 \\
\hline
\end{tabular}
\end{table}

\section{The Financial Structure of Service Contracts}

Service contracts establish a financial relationship between a principal government that pays for a service and an agent that delivers it. In that sense, directionality is not a technical detail but the substantive content of the relationship. A tie from City A to County B means something categorically different from an undirected tie: one government is purchasing a service, whether because it lacks the capacity to produce it, finds it more efficient to outsource, or faces other fiscal or organizational constraints, and the other is delivering it under defined terms of accountability. Which governments consistently buy services and which consistently deliver them is not recoverable from a symmetric representation, despite its importance for understanding how local governments divide service responsibilities across jurisdictions.

We focus on service contracts, because they are the only institutional form with a clear financial principal-agent structure. To identify them across the full corpus of 21,629 agreements, we apply the classification model described in \ref{sec:classifier}, which achieves a weighted F1-score of 0.92 on this class. To validate the automated extraction of contracting parties and financial values, we drew a random sample of 100 service contracts and had our labeling team manually verify each case against the source document; 96\% of extractions were confirmed correct. The network construction and entity disambiguation procedures are described below; what follows reports what the network reveals.

\subsection{Financial Contracting Patterns by Type of Local Government}
\begin{table*}[ht]
\centering
\caption{Financial contracting patterns by principal and agent organization type, computed over all service-contract edges in the network ($N = 2,726$) after only including County, City and Special districts as principals. For each principal type (County, City, Special District), rows show every agent type that principal contracts with. N is the number of contracts for that principal--agent pair; \% is that pair's share of the principal type's total contract count (i.e., \% within a principal group sums to 100\%). Avg Amount is the mean dollar value per contract for that pair. Top Services lists the four service categories occurring most frequently for that pair, ordered from most to least frequent.}
\label{tab:principal_agent}
\small
\begin{tabular}{llrrrp{4.5cm}}
\toprule
\textbf{Principal} & \textbf{Agent} & \textbf{N} & \textbf{\%} & \textbf{Avg Amount} & \textbf{Top Services} \\
\midrule
County & County & 264 & 37.60\% & \$200,968 & Street and Road Systems, Health, Engineering, Jail and Corrections \\
       & City & 265 & 37.70\% & \$235,860 & Street and Road Systems, Emergency Management, Fire Response, Hazmat Response \\
       & State Agency & 75 & 10.70\% & \$111,127 & Jail and Corrections, Health, Motor Vehicles, Information Services \\
       & Regional & 33 & 4.70\% & \$132,740 & Jail and Corrections, Health, Other Public Works, Planning \\
       & Special District & 51 & 7.30\% & \$33,575 & Fire Response, Other Public Works, Education, Emergency Management \\
       & Other & 14 & 2.00\% & \$376,112 & Information Services, Street and Road Systems, Health, Jail and Corrections \\
\midrule
City & County & 1577 & 84.50\% & \$206,029 & Police Protection, Street and Road Systems, Other Public Works, Health \\
     & City & 173 & 9.30\% & \$184,442 & Police Protection, Emergency Management, Fire Response, Street and Road Systems \\
     & State Agency & 42 & 2.30\% & \$181,598 & Jail and Corrections, Criminal Investigation, Health, Other Public Works \\
     & Regional & 21 & 1.10\% & \$422,717 & Public Transit, Other Public Works, Sanitation, Planning \\
     & Special District & 47 & 2.50\% & \$2,652,138 & Education, Fire Response, Parks and Recreation, Facilities \\
     & Other & 6 & 0.30\% & \$954,199 & Emergency Management, Parks and Recreation, Facilities, Information Services \\
\midrule
Special District & County & 32 & 20.30\% & \$160,496 & Education, Police Protection, Street and Road Systems, Other Public Works \\
                 & City & 92 & 58.20\% & \$167,187 & Education, Fire Response, Police Protection, Parks and Recreation \\
                 & State Agency & 7 & 4.40\% & \$60,887 & Education, Jail and Corrections, Information Services, Community and Neighborhood \\
                 & Regional & 3 & 1.90\% & \$15,860 & Information Services, Motor Vehicles \\
                 & Special District & 24 & 15.20\% & \$15,099 & Education, Police Protection, Facilities, Motor Vehicles \\
\bottomrule
\end{tabular}
\parbox{\linewidth}{\small\vspace{4pt}
}
\end{table*}

Table ~\ref{tab:principal_agent} presents the distribution of financial contracting relationships by principal and agent type. Regarding general purpose governments, cities contracting with counties represent the most concentrated dyad in the network, with 84.5\% of city contracts directed to county governments, predominantly for police protection and street and road systems. This pattern reflects cities purchasing public safety and infrastructure capacity from counties, whether because they lack the capacity to produce these services, find it more efficient to outsource, or face other fiscal or organizational constraints, with counties serving as the primary service providers \citep{li2021local, zhang2024exploring}. City-to-city contracting, by contrast, is far less common (9.3\%) and oriented toward emergency management and fire response, suggesting that when cities contract with each other it is primarily for services where operational capacity is distributed across neighboring jurisdictions rather than consolidated in a single provider.

The city-to-special district relationship stands out for a different reason. While it represents only 2.5\% of city contracts by volume, it carries the highest average contract value in the network (\$2,652,138), driven by education contracts between municipalities and school districts. The scale of these arrangements reflects the long-term financial commitments involved when cities fund educational services across jurisdictional boundaries. 

County principals distribute contracts more evenly, with roughly equal shares going to other counties (37.6\%) and cities (37.7\%), both concentrated in street and road systems, health, and emergency management. This more balanced distribution reflects the regional coordination role counties play across service domains, where they serve simultaneously as buyers of specialized capacity and as the primary providers of county-wide services to municipalities that cannot sustain them independently.

 Regarding special purpose governments, special district contracting is modest in both scale and scope. School districts and fire districts contract primarily with cities (58.2\%), counties (20.3\%), and other special districts (15.2\%), concentrating their agreements in education and fire response, consistent with the narrow service mandates these governments carry. Special districts are active participants in the contracting network, but they operate within a constrained range, buying and selling within the boundaries of what their enabling legislation permits.

\subsection{The core contracting network}

Then, we built a weighted, directed, contracting network, where the nodes are the government agencies, edges are directed from principal (payer) to agent (payee), to show who is paying whom. The edge weight is the dollar amount of the contract as the primary signals. Edges also have service area, year filed, and contract duration where available, as attributes.

\begin{figure*}[!h]
    \centering
    \includegraphics[width=\textwidth]{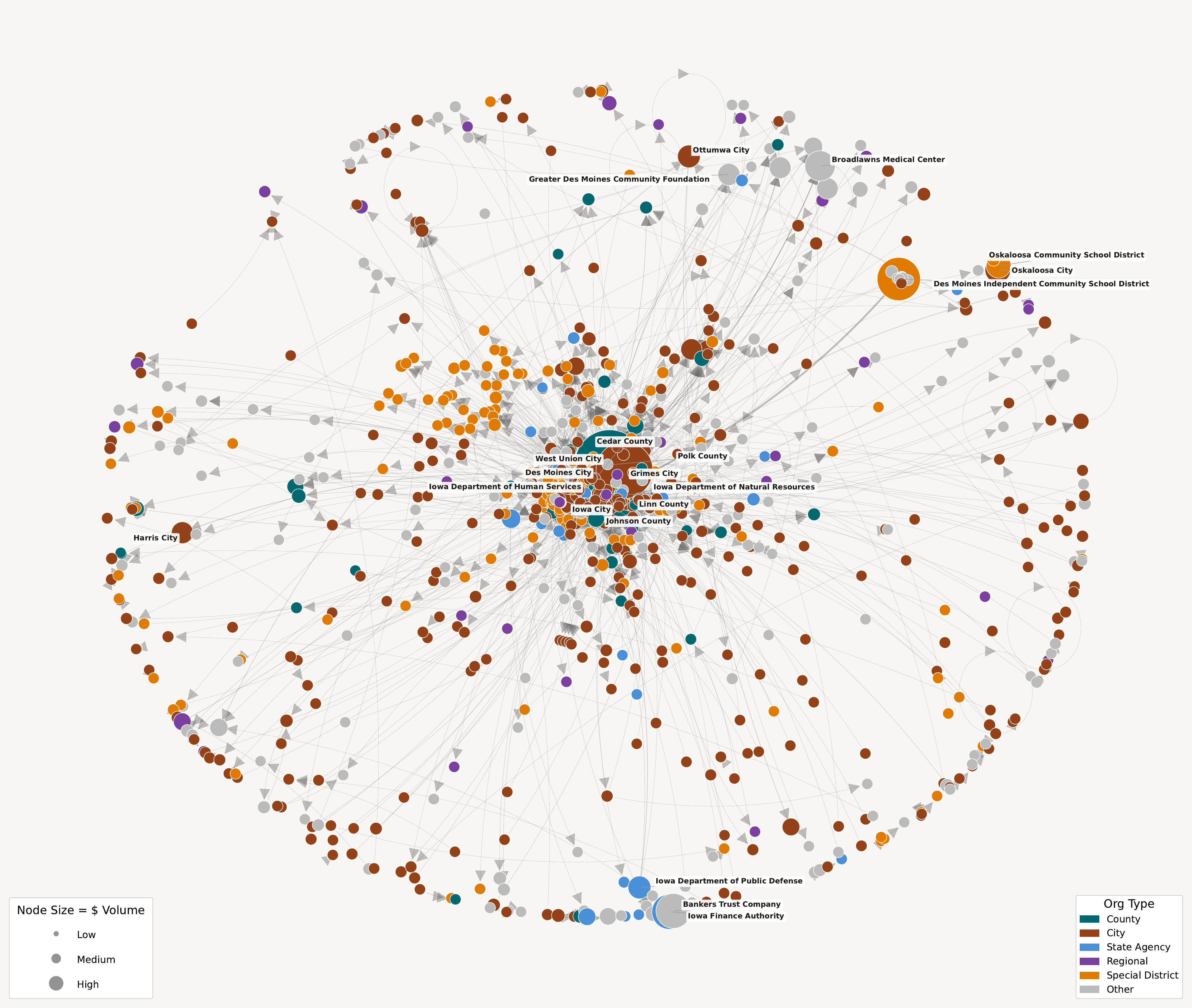} 
    \caption{Directed financial network of Iowa 28E service contracts with a transaction value of at least \$1{,}000, restricted to service contracts. Nodes represent contracting agencies; an arrow from one node to another indicates a payment relationship, pointing from the paying agency (principal) to the receiving agency (agent). Node color indicates organization type; node size is proportional to each agency's total dollar volume (sum of incoming and outgoing payments). Of 3{,}234 total service-contract edges, 1{,}420 (44\%) exceed the \$1{,}000 threshold and are shown, representing 99\% of total dollar volume in the corpus. For visibility, only the 20 highest-volume nodes are labeled.}
    \label{fig:network} 
\end{figure*}

Figure~\ref{fig:network} visualizes the network, edge thickness proportional to dollar amount. The network only shows transactions above \$1000, and only the top 20 actors have labels.

\begin{figure*}[!h]
    \centering
    \includegraphics[width=\textwidth]{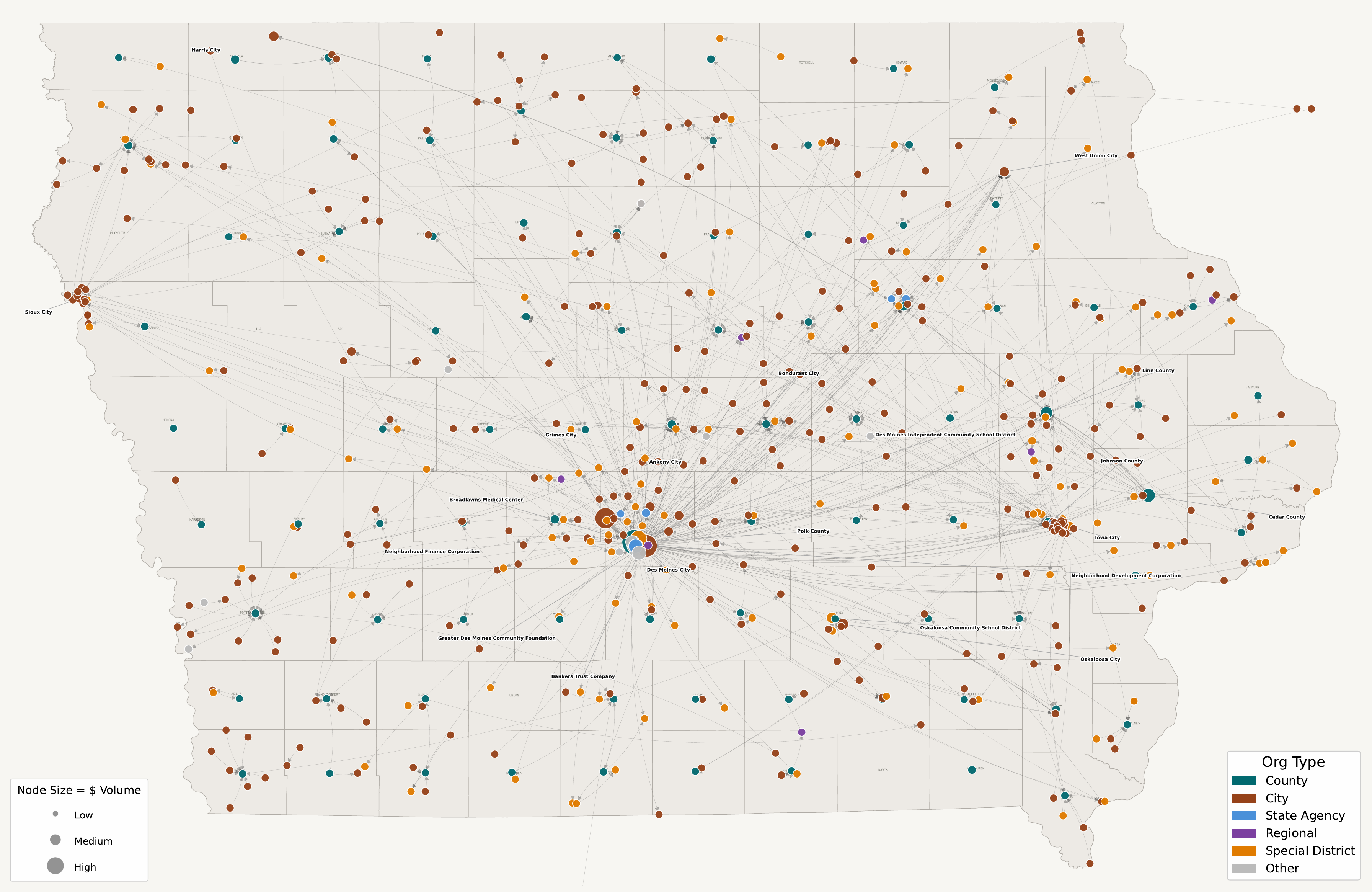} 
    \caption{Geographic layout of the Iowa 28E financial network, restricted to service contracts with a transaction value of at least \$10{,}000. Nodes represent contracting agencies, plotted at their approximate geographic location over an Iowa county basemap; edges connect agencies with a qualifying payment relationship between them. Node color indicates organization type; node size is proportional to each agency's total dollar volume (Low/Medium/High, per legend), summing incoming and outgoing payments. Of 3{,}234 total service-contract edges, 734 (22\%) exceed the \$10{,}000 threshold and are shown. Only the 20 highest-volume nodes are labeled.}
    \label{fig:network_map} 
\end{figure*}

To find the most influential transactions, we filtered out the transactions less than $\$10,000$ between principal and agents, and put them on the map of Iowa. As expected the densest cluster is around the Des Moines City. Figure~\ref{fig:network_map} shows this map, with 590 nodes and 734 edges.

\subsection{Descriptive Network Analysis}

Table~\ref{tab:network_summary} summarizes the basic structural properties of the contracting network. The network consists of 1,642 agencies connected by 3,234 directed financial ties, with a density of 0.0012, suggesting that intergovernmental contracting for service delivery in Iowa is sparse rather than a single integrated market. Most governments contract with only a narrow set of partners, and the network as a whole reflects the fragmented structure of local governance in Iowa rather than a system of broadly interconnected fiscal relationships.

Reciprocity is 0.160, suggesting that roughly 16\% of financial ties have a return tie in the opposite direction, indicating that some actors occupy both principal and agent roles in relation to the same partner across different service areas. This is a meaningful share. It tells us that the principal-agent distinction, while structurally dominant, is not absolute. Some governments buy certain services from a partner while simultaneously selling others back to that very same partner, a pattern consistent with the regional interdependence that characterizes interlocal contracting in states with highly fragmented local government systems.

Transitivity is low (0.0183), indicating that financial contracting relationships do not cluster into closed triads. Governments that share a common contractor rarely contract with each other directly. This likely reflects the asymmetric and service-specific nature of these arrangements: contracting relationships are organized around what each government needs and what each can provide, not around social proximity or generalized exchange. The result is a network organized around a limited number of service-producing governments that maintain contractual relationships with multiple buyers that are otherwise unconnected to one another.

\begin{table}[H]
\centering
\caption{Network Summary Statistics}
\label{tab:network_summary}
\begin{tabular}{lr}
\toprule
\textbf{Property} & \textbf{Value} \\
\midrule
Nodes                    & 1,642         \\
Directed edges        & 3,234         \\
Network density                    & 0.0012        \\
Reciprocity                        & 0.1600        \\
Transitivity                       & 0.0183        \\
Out-degree centralization          & 0.1080        \\
Betweenness centralization         & 0.2492        \\
\bottomrule
\end{tabular}
\end{table}

Regarding the global measures of network centralization, out-degree centralization (0.1080) indicates moderate concentration in purchasing activity, with a handful of actors initiating the majority of financial contracts. This is consistent with the dyadic patterns reported above, where a small number of county and city governments account for a disproportionate share of contracting volume. Betweenness centralization (0.2492) is notably higher, and the gap between the two centralization measures is informative. It means that the governments driving the most resource flow are not necessarily the most active buyers, but rather those that sit between otherwise disconnected parts of the network. A government with high betweenness and modest out-degree is one that connects clusters of governments that would otherwise have no financial relationship, a structurally critical position that purchasing volume alone does not capture.

Overall, the network resembles a fragmented system of specialized exchange rather than a densely interconnected structure of mutual coordination. A relatively small number of governments occupy central service-delivery positions, while most contractual relationships remain narrow, asymmetric, and functionally specific.

\subsection{Null Configuration Model}

\begin{figure*}[!h]
    \centering
    \includegraphics[width=\textwidth]{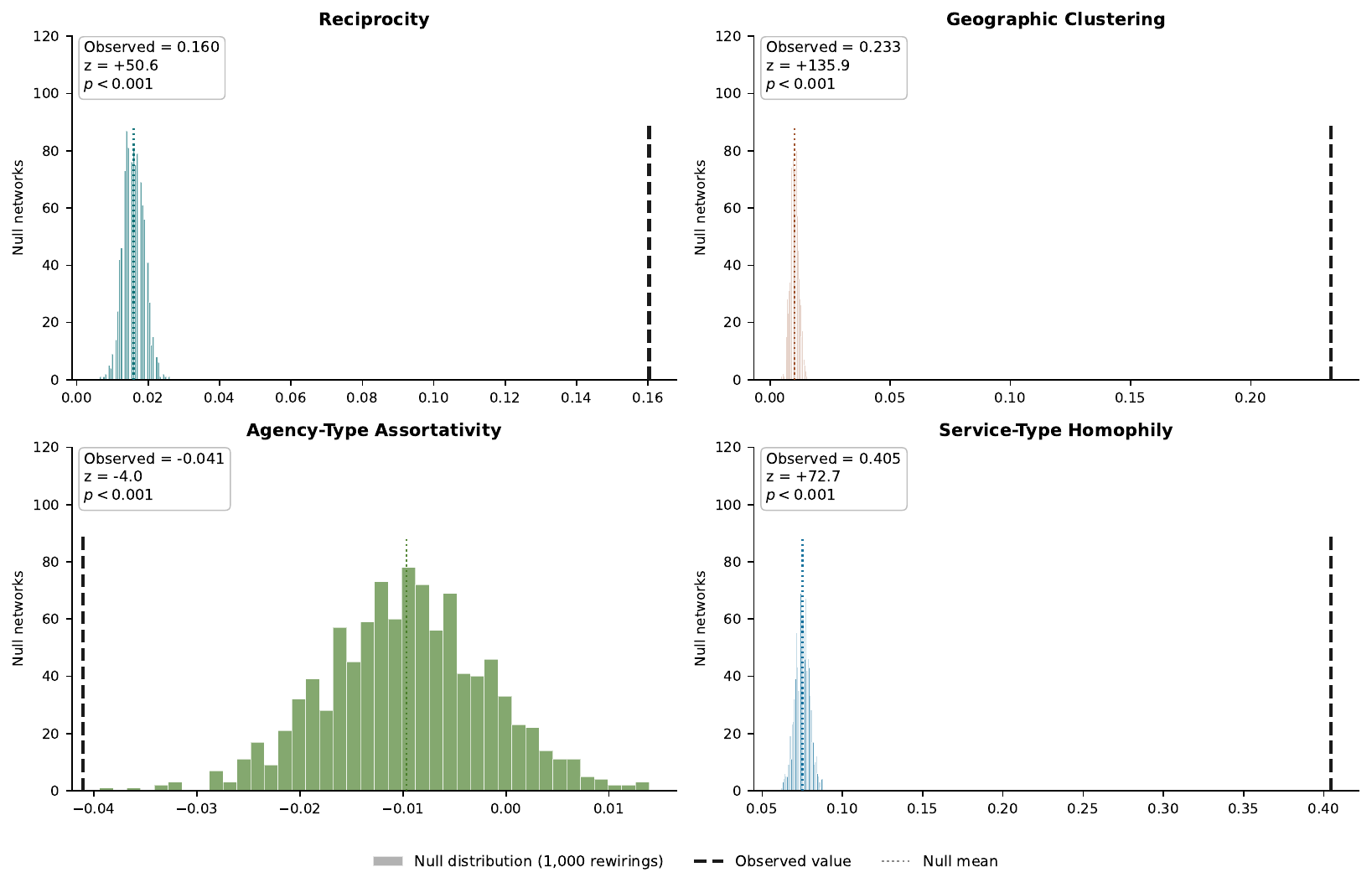}
    \caption{Four network properties compared against a configuration-model null distribution. For each property, the dashed vertical line marks the value observed in the real network; the histogram shows the distribution of that same property computed on 1{,}000 randomly rewired versions of the network that preserve each node's in-degree and out-degree (number of payment partners) but randomize which agencies are actually connected. The dotted line marks the mean of the null distribution. Reciprocity is the share of directed ties that are reciprocated by a tie in the opposite direction. Geographic Clustering is the share of ties connecting two agencies in the same county. Agency-Type Assortativity is Newman's assortativity coefficient (ranging from $-1$ to $1$)~\citep{newman2003mixing} computed on agency organization type, with positive values indicating agencies preferentially contract with same-type agencies. Service-Type Homophily is the share of ties connecting two agencies whose dominant service category matches. Each panel reports a two-sided empirical p-value, the share of the null draws at least as far from the null mean as the observed value in either direction, with the $(k+1)/(N+1)$ resampling correction. We also report z-score, the observed value minus the null mean in units of the null standard deviation, to distinguish the panels. All four observed values fall outside the range of all 1,000 rewirings.}
    \label{fig:null_model}
\end{figure*}

Figure~\ref{fig:null_model} compares each observed value (dashed line) against the range produced by 1{,}000 randomly rewired networks (shaded histograms), which keep each agency's number of partners but scramble who contracts with whom. A property is meaningful only if the observed value falls well outside this random range. We use two-sided empirical p-values with $(k+1)/(N+1)$ resampling correction; no rewiring matched the observations, so all four report $p-value < 0.001$. To distinguish the identical p-values, we also report z-scores for each panel. Z-score is the distance from the null mean in null standard deviations, as a descriptive measure of how outside the null range a value falls.

All four properties fall far outside the random range, though two do so in the direction opposite our expectation. Agencies contract with partners in their own county and within their own service area far more than their number of connections alone would predict (geographic clustering $0.23$ and service-type homophily $0.41$, versus random averages near $0.01$ and $0.08$; $z=+135.9$ and $z=+72.7$, both $p < 0.001$). Reciprocity, which principal--agent theory led us to expect below chance, is instead well above it ($0.16$ versus $0.02$; $z=+51$, $p<0.001$), confirming that the two-way ties discussed earlier are a genuine structural feature rather than an artifact of a few high-volume hubs. Assortativity by agency type likewise runs opposite to expectation: rather than cities preferentially contracting with cities, the observed value ($-0.04$) is more negative than any rewired network produced, meaning money crosses organizational types even more than chance would predict, though this is the weakest of the four effects ($z = -4.0)$, against a wide null).

\subsection{Weighted Hierarchical Stochastic Block Model (SBM)}

The hierarchical degree-corrected weighted SBM finds the latent block structure that most parsimoniously explains the observed pattern of financial flows. It uses Peixoto's~\citep{peixoto2014hierarchical} minimum description length criterion for model selection with no need to pre-specify the number of blocks, using the graph-tool~\citep{peixoto_graph-tool_2014} for Python.
Intuitively, the model sorts agencies into groups, or blocks, so that agencies in the same block send and receive money to and from similar partners in similar amounts, meaning that they occupy equivalent positions in the flow of funds, regardless of who they are or where they sit. Two features matter for our data. First, the model is `degree-corrected': because a handful of agencies are far more active than the rest, and is a heavy-tailed distribution. Second, it chooses the number of blocks automatically using a `minimum-description-length' criterion. The structure we report is therefore the most parsimonious one the data support, not a number we fixed in advance. The model returns a nested hierarchy.

Table~\ref{tab:blocks} and Figure~\ref{fig:flow_matrix} describe the eight base blocks and the money flowing between them. The clearest organizing principle is financial role: four blocks are net payers `principals' and four are net recipients `agents'. Block membership does not follow government type or location. The smallest block, just 83 agencies, receives the largest total inflow of any group, a few large providers absorbing much of the contracting revenue. The flow matrix tells the same story from the other direction: money is not confined within blocks but moves substantially across them, indicating an integrated system rather than a set of separate clusters.

This role-based structure reconciles two earlier findings. Individual ties are strongly localized: agencies contract mostly within their own county and service area, yet the blocks themselves are geographically and organizationally mixed. 

\begin{table*}[ht]
\centering
\caption{Characteristics of the eight blocks identified by the stochastic block model (SBM) fit on the full Iowa 28E service-contract network. Blocks are labeled by their arbitrary model-assigned integer ID, which carries no inherent order or meaning. Agencies is the number of agencies belonging to each block. Out (\$M) and In (\$M) are, respectively, the total dollars sent and received by all agencies in that block, summed across all their contracts, in millions of dollars. Net (\$M) is Out minus In. Role classifies a block as \textit{Principal} if Net is positive (the block is a net payer overall) or \textit{Agent} if Net is negative (a net recipient).}
\label{tab:blocks}
\renewcommand{\arraystretch}{1.2}
\begin{tabular}{lrrrrl}
\hline
Block & Agencies & Out (\$M) & In (\$M) & Net (\$M) & Role \\
\hline
93  & 83  & 338.8 & 538.7 & $-199.9$ & Agent \\
40  & 211 & 257.9 & 13.0  & 244.9    & Principal \\
251 & 131 & 168.8 & 166.7 & 2.1      & Principal \\
110 & 258 & 95.4  & 133.7 & $-38.3$  & Agent \\
883 & 234 & 71.5  & 94.1  & $-22.6$  & Agent \\
995 & 318 & 62.2  & 27.5  & 34.7     & Principal \\
2   & 273 & 17.6  & 7.1   & 10.4     & Principal \\
842 & 134 & 7.9   & 39.3  & $-31.4$  & Agent \\
\hline
\end{tabular}
\end{table*}

\begin{figure}[t]
    \centering
    \includegraphics[width=\columnwidth]{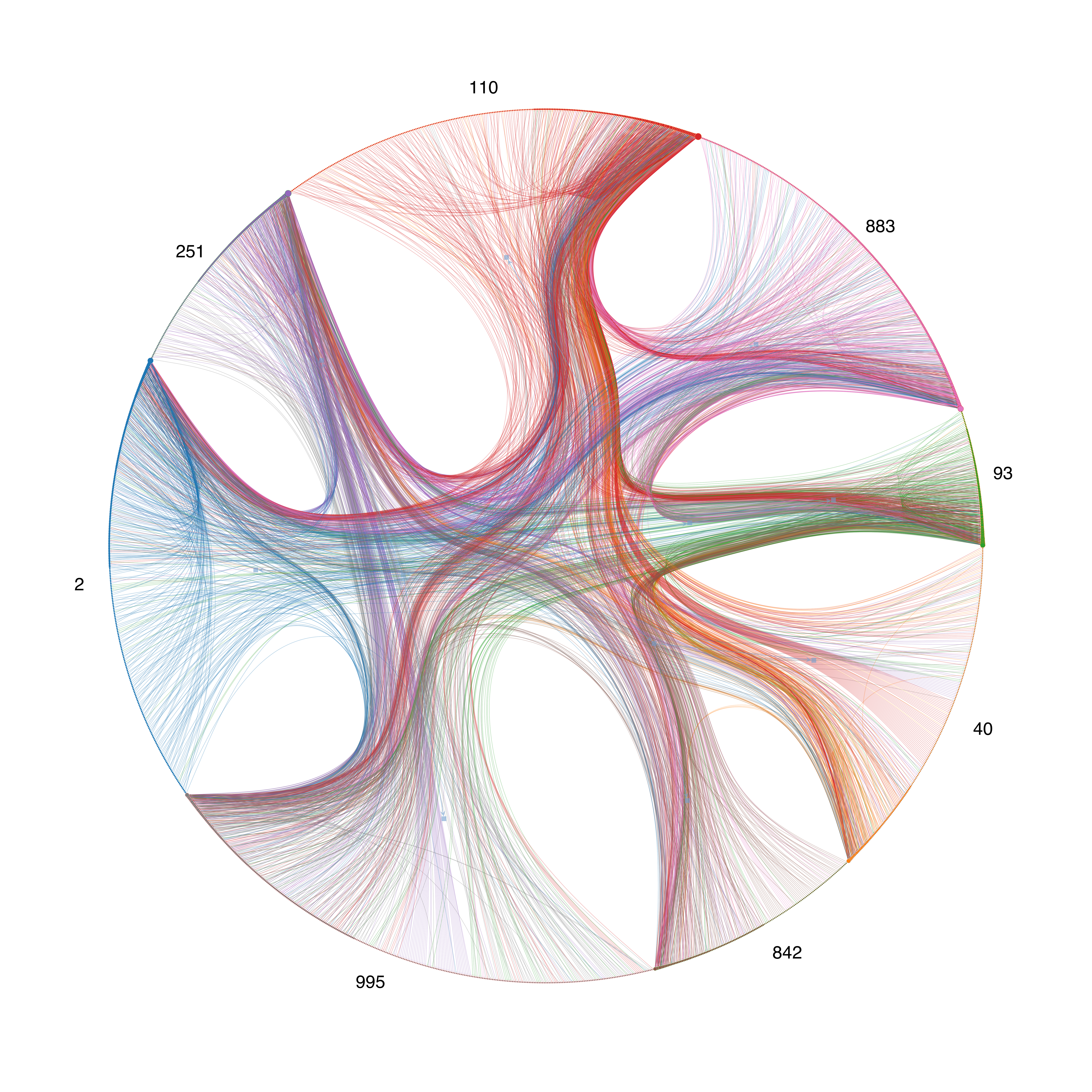}
    \caption{Nested stochastic block model (SBM) of the Iowa 28E financial network. The model groups agencies into blocks such that agencies in the same block occupy structurally equivalent positions in the flow of funds. This figure displays the eight base-level blocks identified by the model. Each of the eight colored arc segments around the circle represents one block, grouping all member agencies along that segment and labeled with its model-assigned block ID; these IDs match those in Table~\ref{tab:blocks} and Figure~\ref{fig:flow_matrix}. Curved lines connect individual agencies across different blocks and represent directed payment relationships between them, colored by their block of origin. The dense crossing of lines through the center shows that money moves substantially between blocks rather than remaining within them (quantified in Figure~\ref{fig:flow_matrix}).}
    \label{fig:sbm_hierarchy}
\end{figure}

The outcome of the model groups agencies into eight blocks at the finest level. These eight blocks in turn combine into four broader groups, and above that all agencies merge into one. Figure~\ref{fig:sbm_hierarchy} displays this structure as a circle, where each colored segment is one block and the curved lines are the financial ties running between agencies in different blocks; the dense web of
cross-colored lines through the center shows that money moves freely between blocks rather than staying within them.

\begin{figure}[t]
    \centering
    \includegraphics[width=\columnwidth]{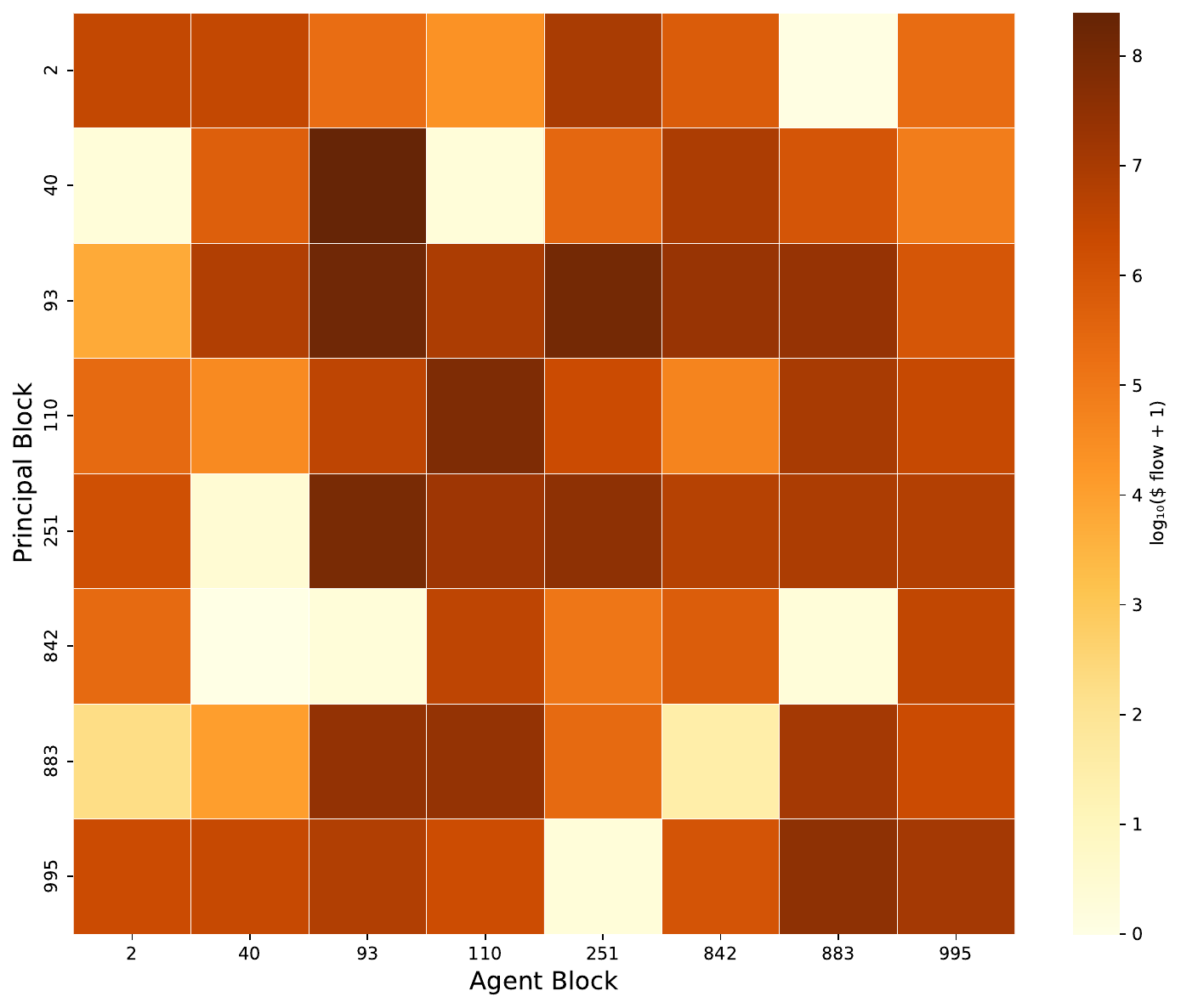}
    \caption{Total dollar flows between the eight blocks identified by the stochastic block model (SBM), aggregated over all service contracts in the network. Rows are paying (principal) blocks; columns are receiving (agent) blocks. Cell color encodes total dollar flow from the row's block to the column's block, on a $\log_{10}(\$ \text{ flow} + 1)$ scale (colorbar). The $+1$ offset avoids undefined values for block pairs with zero flow. Block identifiers on both axes (2, 40, 93, 110, 251, 842, 883, 995) match the arbitrary labels assigned by the SBM in Table~\ref{tab:blocks}, and are not ordinal. Diagonal cells show flow within a block (both parties belonging to the same block); off-diagonal intensity across nearly the entire matrix shows that most dollar volume moves between blocks rather than staying within them.}
\label{fig:flow_matrix}
    \label{fig:flow_matrix}
\end{figure}

\section{Discussion: Broader Implications and Future Directions}
The findings reported here carry implications that extend beyond the methodological contribution. 
Studies of interlocal collaboration have made significant progress in understanding why governments collaborate and with whom, but the institutional content of the agreements themselves has remained difficult to analyze at scale. Manual coding is costly and cannot cover archives of tens of thousands of documents. The pipeline introduced here addresses that constraint directly, making it possible to classify the full Iowa 28E archive by institutional form and map the financial structure of service contracts across nearly three decades of contracting activity.

The network analysis reveals that intergovernmental service delivery in Iowa is not a web of mutual exchange but a stratified system with clear positional differences across government types. Counties occupy the most central structural positions, functioning simultaneously as buyers of specialized capacity and as the primary service providers for municipalities that cannot sustain certain functions independently. Cities are predominantly buyers, dependent on counties for public safety and infrastructure in ways that are not visible from collaboration counts alone. Special districts operate at the margins, constrained by narrow service mandates and dependent on general-purpose governments for functions outside those mandates. These positional differences have direct implications for how accountability is distributed across the intergovernmental system, and they are visible here at a scale that prior work has not been able to achieve.

For public administration scholars, the value of this approach is not that it asks new questions but that it makes existing ones answerable with data of a scope and completeness that manual coding cannot produce. Questions about which governments are systematically excluded from collaborative arrangements, how financial relationships evolve over time as fiscal conditions change, and whether structural position in the contracting network is associated with service delivery outcomes have been on the field's agenda for some time. What has been missing is the data infrastructure to address them at scale. Iowa's classified archive provides that infrastructure, and extending this pipeline to other states and service domains is the most direct path toward comparative analysis.

For practitioners and public managers, the pipeline offers a practical tool for improving transparency in how governments coordinate service delivery. State oversight agencies can use a classified corpus to identify patterns in how local governments structure collaboration, track how financial relationships evolve, and flag contracting arrangements that warrant closer scrutiny. The publicly available labels, extracted financial data, and codebase are designed to support exactly this kind of use.

For data scientists and computational social scientists, the pipeline illustrates how recent advances in document processing reshape what is feasible with archives that were previously too noisy to analyze. Neural network OCR tools now extract clean text from scanned and photographed records that older systems could not handle reliably, making decades of legacy documents available for quantitative analysis with far less preprocessing noise. On the classification task itself, the results point to a practical division of labor among methods. When no labeled data are available, chain-of-thought prompting of a large language model achieves usable accuracy directly, though at a computational cost that makes it expensive to run across a full
archive; its outputs are trustworthy in some categories but not all, so validation against human judgment remains necessary rather than optional. Once a modest set of labeled documents exists, embeddings become the more efficient path: by translating
words and their context into numerical vectors, they allow lightweight machine learning models to reach high accuracy at a fraction of the cost. The broader lesson is that large language models are best treated not as a single tool but as components
to be combined and verified according to the data available, and that, with the right pipeline, even the network structure recovered here becomes reproducible from raw documents rather than from hand-curated records.

\section{Conclusion}

This paper introduced a computational pipeline for classifying intergovernmental agreements by institutional form and extracting financial relationships between principals and agents in service contracts. Applied to Iowa's 28E archive, the largest and most comprehensive dataset of interlocal agreements available in the United States, it produces the first large-scale classified dataset of its kind and uses it to map how money moves across local governments through formal service arrangements. But our findings carry some uncertainty from the classification and extraction steps that identify service contracts and their financial terms: precision on the service-contract class is 0.89, meaning roughly one in ten agreements flagged as service contracts is a false positive, and manual verification of the extracted principal, agent, and dollar amounts found a 4\% error rate.

The contributions are both methodological and empirical. On the methodological side, the pipeline addresses the core challenges of working with unstructured legal documents at scale, including noise, class imbalance, document length, and small labeled datasets, and makes the classified labels, extracted financial relationships, network data, and pipeline code publicly available for other researchers to build on. On the empirical side, the network analysis reveals a system that is sparse, asymmetric, and organized around a small number of structurally central governments, with counties serving as the most versatile actors, cities as predominantly buyers, and special districts operating at the margins of the contracting network.

Several directions follow naturally from this work. A longitudinal extension would make it possible to track how financial relationships between governments form and shift over time in response to fiscal conditions and service demands, questions the field has asked but that data constraints have limited. Extending the pipeline to states with partial filing records, and developing methods for handling incomplete archives, would enable comparative analysis across different institutional contexts. And the classified corpus itself opens a range of questions about why governments choose particular collaborative arrangements that descriptive analysis alone cannot answer, but that prior theoretical work in the field has laid the groundwork to pursue. The classified dataset, extracted financial relationships, network data, and pipeline code are publicly available at \url{https://anonymous.4open.science/r/iowa-28e-pipeline-85AF}.

\section*{Declaration of generative AI}

Statement: During the preparation of this work some of the authors used LLM powered technologies in order to review the paper, grammar and spell checking. After using this tool/service, the author(s reviewed and edited the content as needed and take full responsibility for the content of the published article.

\section*{Acknowledgment}

We are grateful to the Networks and Governance Lab at the University of Illinois Chicago (UIC) for collecting, hand-coding, and sharing the initial training dataset of classified interlocal agreements that formed the foundation for this pipeline. We appreciate Julia Zimmerman, Ben Cooley, and Laurent Hébert-Dufresne for their valuable thoughts shaping the graphs. We also thank the Complex Networks Winter workshop where we first started this project. We thank Brennan Klein, Keiko Nomura, and Laura Vander Meiden for feedback on earlier versions of this project. Our team acknowledges support from the MassMutual Center of Excellence in Complex Systems and Data Science (Grant \# FP2860), Alfred P. Sloan Foundation (Grant \#G-2024-22498), and the National Science Foundation (Grant \#2242829).

\bibliographystyle{apacite} 
\bibliography{references_apa}

\appendix

\section{Hyper Parameters}
\label{app:training_details}

To train the ML classifiers, we used the below hyperparameters for gird search to be able to find the most accurate classifier.

\begin{lstlisting}[style=configstyle, language=Python, 
    caption={Hyperparameter search space and cross-validation configuration.},
    label={lst:param_sweep}]
cv = StratifiedKFold(n_splits=5, shuffle=True)

models = [
    # Logistic Regression
    ("logreg", Pipeline([
        ("scale", StandardScaler()),
        ("pca",   "passthrough"),
        ("clf",   LogisticRegression(max_iter=5000,
                  class_weight="balanced"))]),
     {"pca":   ["passthrough", PCA(n_components=0.95)],
      "clf__C": [0.1, 1, 10]}),

    # Linear SVM
    ("linear_svm", Pipeline([
        ("scale", StandardScaler()),
        ("pca",   "passthrough"),
        ("clf",   LinearSVC(class_weight="balanced"))]),
     {"pca":    ["passthrough", PCA(n_components=0.95)],
      "clf__C": [0.1, 1, 10]}),

    # RBF SVM
    ("rbf_svm", Pipeline([
        ("scale", StandardScaler()),
        ("pca",   "passthrough"),
        ("clf",   SVC(kernel="rbf", class_weight="balanced"))]),
     {"pca":       ["passthrough", PCA(n_components=0.95)],
      "clf__C":     [1, 10, 100],
      "clf__gamma": ["scale", 1e-2, 1e-1]}),

    # Poly SVM
    ("rbf_svm", Pipeline([
        ("scale", StandardScaler()),
        ("pca",   "passthrough"),
        ("clf",   SVC(kernel="poly", class_weight="balanced"))]),
     {"pca":       ["passthrough", PCA(n_components=0.95)],
      "clf__C":     [1, 10, 100],
      "clf__gamma": ["scale", 1e-2, 1e-1]}),
      
    # Ridge Classifier
    ("ridge", Pipeline([
        ("scale", StandardScaler()),
        ("pca",   "passthrough"),
        ("clf",   
        RidgeClassifier(class_weight="balanced"))]),
     {"pca":       ["passthrough", PCA(n_components=0.95)],
      "clf__alpha": [0.1, 1, 10]}),

    # Extra Trees
    ("trees",
     ExtraTreesClassifier(n_estimators=1000, n_jobs=-1,
                          class_weight="balanced"),
     {"max_depth":        [None, 20],
      "min_samples_leaf": [1, 2],
      "max_features":     ["sqrt", 0.5]}),

    # MLP
    ("mlp", Pipeline([
        ("scale", StandardScaler()),
        ("pca",   "passthrough"),
        ("clf",   MLPClassifier(max_iter=2000,
                    early_stopping=True))]),
     {"pca":                  ["passthrough", PCA(n_components=0.95)],
      "clf__hidden_layer_sizes": [(128,), (256,)],
      "clf__alpha":              [1e-4, 1e-3]}),
]
\end{lstlisting}

\section{Pipeline Schematic}
\label{app:pipeline}

\begin{figure*}[p]
    \centering
    \includegraphics[width=\textwidth]{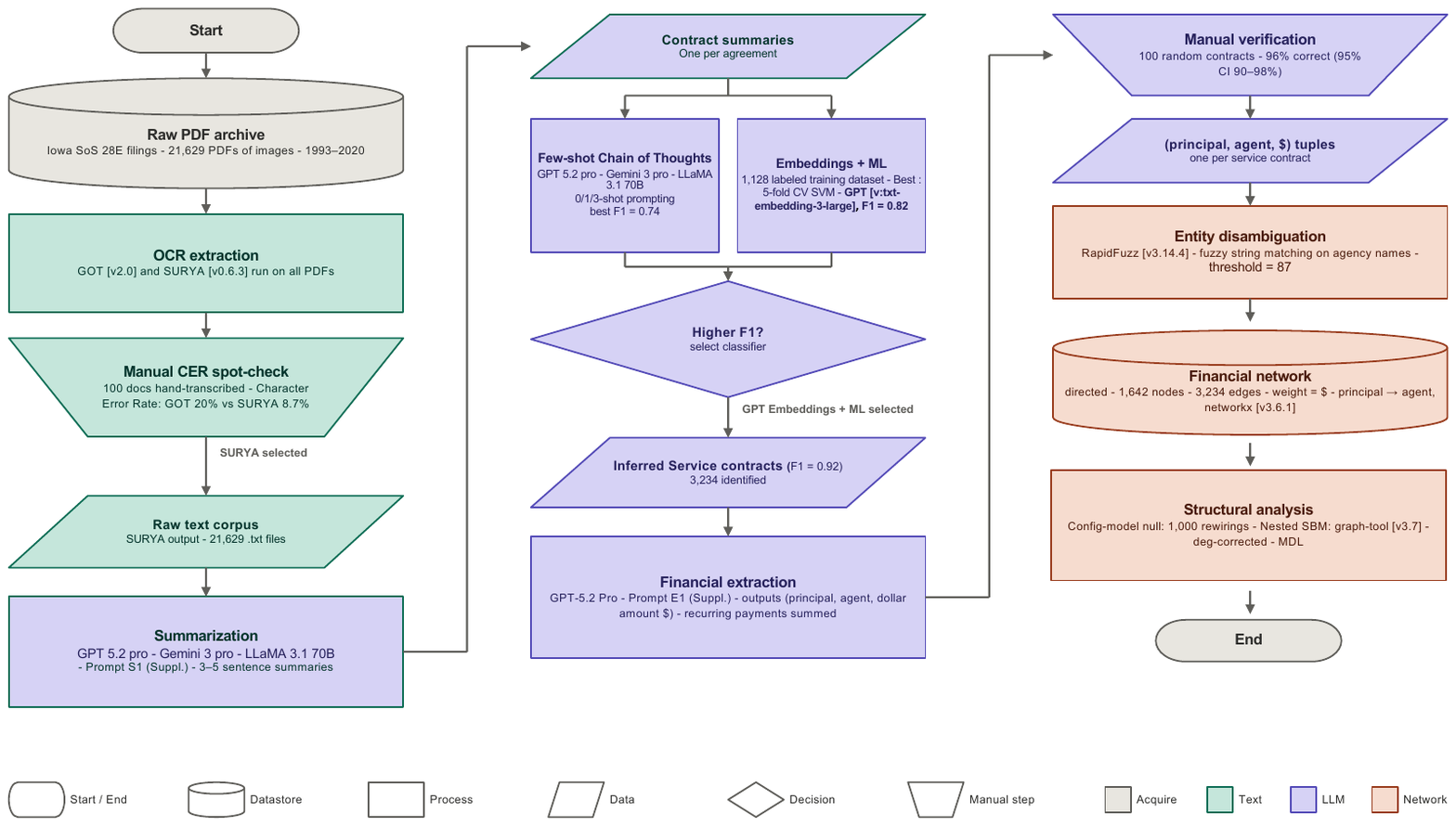}
    \caption{Complete processing pipeline from raw PDF archive to structural network analysis. Each automated step names the specific tool or model used; decision points show the selection criterion and outcome (OCR engine selected by Character Error Rate; classification approach selected by weighted F1); manual-operation shapes mark the two human validation checkpoints (the 100 document CER spot check and the 100 contract extraction verification). Parallelograms show the data
    representation flowing between stages.}
    \label{fig:pipeline_flowchart}
\end{figure*}

\begin{figure*}[p]
    \centering
    \includegraphics[width=\textwidth]{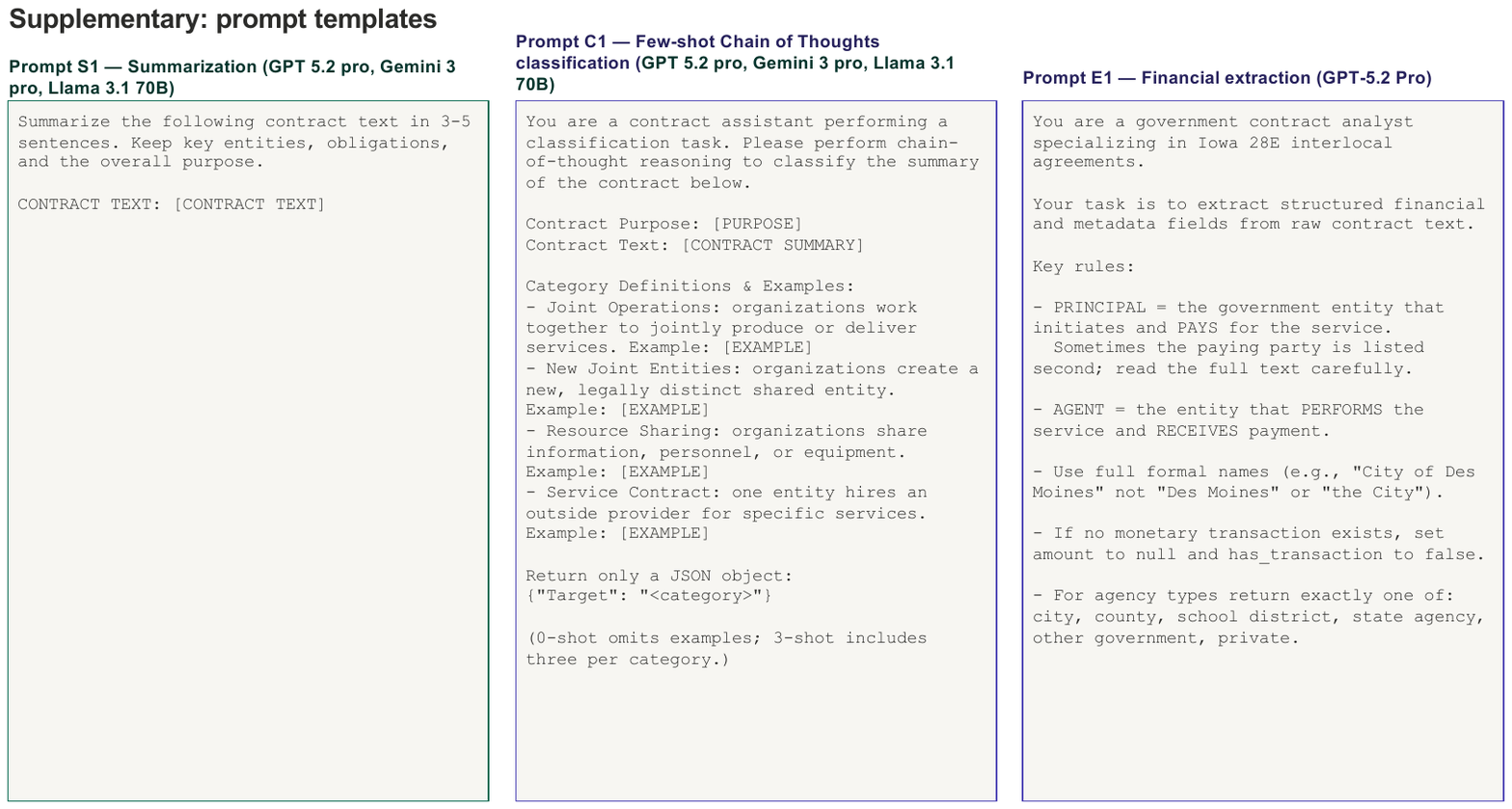}
    \caption{Prompt templates used at each LLM calling stage: S1 (contract summarization), C1 (few-shot chain-of-thought classification, shown in one-shot form), and E1 (financial entity extraction).}
    \label{fig:prompt_templates}
\end{figure*}

Figure~\ref{fig:pipeline_flowchart} documents the complete processing pipeline, from the raw PDF archive to the final network analysis. The diagram follows standard flowchart conventions. Colors group stages by phase: data acquisition, text processing, LLM tasks, and network analysis. The prompt templates referenced in the diagram (S1, C1, E1) are provided in Figure~\ref{fig:prompt_templates}.

\end{document}